\documentclass[a4 paper, 10 pt] {article}

\begin{document}

\title{\bf{Infinitely many conservation laws in self-dual Yang--Mills theory}}

\author{C. Adam $^{a)}$\thanks{adam@fpaxp1.usc.es},
J. S\'{a}nchez-Guill\'{e}n $^{a)}$\thanks{joaquin@fpaxp1.usc.es} and
A. Wereszczy\'{n}ski $^{b)}$
\thanks{wereszcz@alphas.if.uj.edu.pl}
\\ \\
$^{a)}$ \small{ Departamento de Fisica de Particulas, Universidad
       de Santiago}  \small{ and Instituto Galego de Fisica de Altas
       Enerxias}\\
       \small{ (IGFAE) E-15782 Santiago de Compostela, Spain}
\\
       \\ $^{b)}$ \small{ Institute of Physics,  Jagiellonian
       University,}
       \small{ Reymonta 4, Krakow, Poland}
 }

\maketitle
\begin{abstract}
Using a nonlocal field transformation for the gauge field known as
Cho--Faddeev--Niemi--Shabanov decomposition as well as ideas
taken from generalized integrability, we derive a new family of
infinitely many conserved currents in the self-dual sector of
$SU(2)$ Yang-Mills theory. These currents may be related to the area
preserving diffeomorphisms on the reduced target space. 
The calculations are performed in a completely covariant
manner and, therefore, can be applied to the self-dual equations in
any space-time dimension with arbitrary signature.
\end{abstract}



\section{Introduction}
A powerful tool in the theory of topological solitons is the
derivation of lower bounds for the energy (or Euclidean action) in
terms of topological charges. Together with these bounds, in some
cases one may derive first order equations (so-called Bogomolny
equations) such that any field configuration obeying these Bogomolny
equations automatically saturates the topological lower bound and is
a true minimizer of the energy (or Euclidean action) functional.
Obviously, any field configuration obeying the Bogomolny equations
automatically obeys the original second order Euler--Lagrange
equations, whereas the converse is not true in general.

In addition to providing true minimizers of the energy functional,
these Bogomolny equations, due to their more restrictive nature,
tend to enhance the number of symmetries and conservation laws.
Sometimes, there exist infinitely many symmetries and infinitely
many conservation laws for the Bogomolny equations. Further, the
Bogomolny equations are usually not of the Euler--Lagrange type,
therefore for those symmetries which are not symmetries of the
original second order system, the issue of conservation laws has to
be investigated separately, that is, Noether's theorem does not
apply. A theory where this happens is, for instance, the CP(1) model
in 2+1 dimensions. For this theory both the infinitely many
symmetries and the infinitely many conservation laws of the
Bogomolny sector have been calculated, e.g., in \cite{ASG1}, and,
indeed, they turn out to be different. Similar investigations for
gauge theories have been performed recently. In the case of the
Abelian Higgs model, an equivalent pattern has been found, i.e.,
there are infinitely many conserved currents in the Bogomolny
sector, and Noether's theorem does not apply, see \cite{we higgs}. A
slightly different scenario is realized in the Abelian projection of
Yang--Mills dilaton theory. There, too, exists a Bogomolny sector,
but this theory has infinitely many symmetries already on the level
of the Lagrangian, therefore the symmetries and conservation laws
are related by Noether's theorem, see \cite{we dilaton}.

In the case of SU(2) Yang--Mills theory, the solutions which
minimize the Euclidean action functional are known as instantons,
and the Bogomolny type first order equations are the self-duality
equations \cite{BPST} - \cite{shuryak}.  The symmetries of the
self-dual sector of SU(2) Yang--Mills theory have been studied by
various authors \cite{chau1} - \cite{ward}.
The result is that the system posesses infinitely
many symmetries and that almost all of them are nonlocal when
expressed in terms of the original fields. A recent review of this
issue can be found in \cite{Papach1}, to which we refer the reader
for further information and additional references. Conservation laws of
self-dual Yang--Mills theory 
related to the non-local symmetries mentioned above have been studied,
e.g., in \cite{prasad} - \cite{ward}. 

A slightly different, more geometric approach to the self-dual Yang--Mills 
(SDYM) equations focusing directly on their integrability has been initiated
by R. Ward \cite{ward2}. In that approach twistor methods are employed, 
and the use of twistor methods in the investigation of the SDYM 
and their conservation laws has played an important role subsequently
(for some recent results, see \cite{wolf1} - \cite{boels1}). 

Another approach to integrability and conservation laws has been proposed in
\cite{joaquin1}, where a generalized zero curvature representation 
suitable for higher-dimensional field theories was
developed, analogously to
the zero curvature representation of Zakharov and Shabat, which
provides integrable field theories in 1+1 dimensions. 
Among other results, it was demonstrated in that paper that
the SDYM 
permit a generalized zero curvature representation. But still only
finitely many conservation laws have been provided for self-dual
Yang--Mills theory in Ref. \cite{joaquin1}. It is the purpose of
the present paper to further develop the issue of integrability and
conservation laws of the self-dual sector of SU(2) Yang--Mills
theory along these lines. 
We will find another set of infinitely many conservation
laws by explicit construction. The corresponding conserved
currents are nonlocal in terms of the original Yang--Mills field,
but they will be local in terms of a well-known nonlocal field
redefinition which we shall use in the sequel. 
In contrast to the conserved currents found previously, the ones
we shall present below are given by manifestly Lorentz covariant 
expressions and may, therefore, easily be generalized to different
space time metrics and dimensions.
Given the relevance
of self-dual Yang--Mills theories both for strong interaction
physics and in a more mathematical context, we believe that the
discovery of these additional conserved currents is of some interest.

We want to remark that in a recent paper devoted to similar problems
\cite{kneipp}, an investigation of integrability in  the sector of
$Z_N$ string solutions of Yang--Mills theory has been performed.
$Z_N$ string solutions are effectively lower dimensional solutions,
but, nevertheless, they also belong to the self-dual sector.
Further, the integrability of self-dual Yang--Mills theories on
certain four-dimensional product manifolds has been used in
\cite{Pop1}, \cite{Pop2} to demonstrate the integrability of abelian
and nonabelian Higgs models on general Riemannian surfaces.

Our paper is organized as follows. 
In Section 2 we present a brief overview of some 
known results on
the self-dual Yang--Mills (SDYM) 
equations. Specifically, we present infinitely many
nonlocal conserved currents as constructed by Prasad et al and by
Papachristou. This overview shall serve later on to relate our own findings to
these already known results.
In Section 3 we 
recapitulate how the self-dual sector  of SU(2) Yang--Mills theory
may be recast into the form of the generalized zero curvature
representation. In Section 4 we introduce the
Cho--Faddeev--Niemi--Shabanov (CFNS) decomposition of the gauge
field and re-express the self-dual equations using this
decomposition. Next, we write down the currents in terms of the
decomposition fields and prove that they are conserved.
Section 5 contains our conclusions.
In the appendix we display the canonical four momenta and field
equations which we need in the main text.
\section{Some known facts about the SDYM}

\subsection{$J$ formulation of the SDYM}
The self-dual sector of $SU(2)$ Yang-Mills theory in Euclidean
space-time is constituted by gauge fields $A_\mu^a$ satisfying the
following equations
\begin{equation}
F^a_{\mu \nu}=^* F^a_{\mu \nu}, \label{sd def}
\end{equation}
where
\begin{equation}
F^a_{\mu\nu} \equiv \partial_\mu A^a_\nu - \partial_\nu A^a_\mu +
\epsilon^{abc}A^b_\mu A^c_\nu \, ,
 \;\;\; ^*F^a_{\mu \nu} \equiv
\frac{1}{2} \epsilon_{\mu \nu \rho \sigma} F^{\rho \sigma}
\end{equation}
It is convenient to rewrite them as
\begin{equation}
F^a_{yz}=0, \;\;\; F^a_{\bar{y} \bar{z}}=0, \;\;\;
F^a_{y\bar{y}}+F^a_{z\bar{z}}=0, \label{sd def1}
\end{equation}
where the new independent variables are defined as
$$
y=\frac{1}{\sqrt{2}}(x_1+ix_2), \;\;\;
\bar{y}=\frac{1}{\sqrt{2}}(x_1-ix_2),
\;\;\;z=\frac{1}{\sqrt{2}}(x_3-ix_4), \;\;\;
\bar{z}=\frac{1}{\sqrt{2}}(x_3+ix_4).
$$
Defining the self-dual gauge fields as
\begin{equation}
A^a_y = g^{-1}_1 \partial_y g_1,\;\;\; A^a_z = g^{-1}_1 \partial_z
g_1, \;\;\; A^a_{\bar{y}} = g^{-1}_2 \partial_{\bar{y}} g_2, \;\;\;
A^a_{\bar{z}} = g^{-1}_2 \partial_{\bar{z}} g_2
\end{equation}
we identically fulfill the first two equations in (\ref{sd def1}).
Here, $g_1, g_2$ are arbitrary group elements in SU(2). Then the
third expression leads to a nontrivial equation giving an equivalent
formulation of the self-dual equations
\begin{equation}
F[J]\equiv \partial_{\bar{y}} \left( J^{-1} \partial_y J
\right)+\partial_{\bar{z}} \left( J^{-1} \partial_z J \right)=0,
\label{sdym J}
\end{equation}
where
\begin{equation}
J=g_1 g_2^{-1}.
\end{equation}
In other words, solutions of the self-dual sector are defined by Eq.
(\ref{sdym J}).
\subsection{Linear system, B\"{a}cklund transformation and hidden symmetry}
There is a formulation of the SDYM equations in terms of a linear
system \cite{chau1}-\cite{chau3}. Namely, consider an auxiliary
matrix field $\psi$ defined by the following set of equations
\begin{equation}
\partial_{\bar{z}} \psi = \lambda \left( \partial_y \psi +
J^{-1}J_y \psi \right), \;\;\; -\partial_{\bar{y}} \psi = \lambda
\left(
\partial_z \psi + J^{-1}J_z \psi \right). \label{ls}
\end{equation}
In fact, this is just the Lax pair formulation. The SDYM equations
(\ref{sdym J}) are derived as a consistency (integrability)
condition $\psi_{\bar{z} \bar{y}}= \psi_{\bar{y}\bar{z}}.$
\\
Further, it is possible to find the related B\"{a}cklund transformation (BT)
\cite{chau1}. It is given by
\begin{equation}
J^{'-1}J'_y -J^{-1}J_y = \lambda (J^{'-1}J)_{\bar{z}}, \;\;\;\;
J^{'-1}J'_z -J^{-1}J_z = \lambda (J^{'-1}J)_{\bar{y}}. \label{bt}
\end{equation}
If $J$ is a solution of SDYM then $J'$ is a new solution of SDYM. 
Further, this BT is an infinitesimal BT i.e., a new
solution generated by the BT may be found performing an infinitesimal
transformation which leaves the SDYM equation invariant. The
pertinent transformation reads
\begin{equation}
J^{-1}\delta J= -\psi T_a \psi^{-1} \alpha^a \label{hidd sym}
\end{equation}
 where $T_a$ is a basis
of the Lie algebra for the gauge fields and $\alpha^a$ are
infinitesimal parameters. Indeed, if we assume that $J'=J+\delta J$
then we get the BT.
\\
This symmetry transformation gives the following commutator
\begin{equation}
[\delta_{\alpha}, \delta_{\beta}] J= \alpha^a\beta^b C^c_{ab}
\frac{d}{d\lambda} (\lambda \delta_c J), \label{commutator 1}
\end{equation}
where $C^c_{ab}$ are the structure constants of the Lie algebra. If we expand
$\psi=\sum_{n=0}^{\infty} \lambda^n \psi^{(n)}$ then we get the
Kac-Moody algebra
\begin{equation}
[\delta_{\alpha}^{(m)}, \delta_{\beta}^{(n)}] J= \alpha^a\beta^b
C^c_{ab} \lambda \delta_c^{(m+n)} J, \label{km algebra}
\end{equation}
where $\delta^{(n)}$ is defined as $J^{-1}\delta J =
\sum_{n=0}^{\infty} \lambda^n J^{-1} \delta^{(n)}J$ and
$\delta^{(n)}J= -J\sum_{n=0}^{\infty} \psi^{(n)} T \psi^{(m-n)}$.
In this way, one may explain the hidden (infinite) symmetries observed
by L. Dolan \cite{dolan}. It is precisely the symmetry
transformation (\ref{hidd sym}) mentioned above.
\subsection{Nonlocal conservation laws}
The SDYM in $J$ formulation is an Euler--Lagrange system, and 
the Noether theorem applies. Indeed, Eq. (\ref{sdym J}) may be easily
derived from the Euclidean action
\begin{equation}
S=\int d^4 x {\rm Tr} [(J^{-1} \partial^\mu J)(J^{-1} \partial_\mu J)] .
\end{equation} 
Therefore, the derivation of an infinite set of symmetries indicates that there
should exist infinitely many conserved quantities, as is expected in any case
for an integrable system. In fact, several families of nonlocal
conserved currents have been found. All these constructions use in
an essential way the $J$-formulation of the self-dual sector and
therefore are unique for $4$-dimensional Euclidean space-time.
Moreover,  manifest Lorentz covariance is lost since we introduced the complex
variables $y,z$. On the other hand, this formulation of the self-dual
equations possesses the advantage that equation
(\ref{sdym J}) has the form of a conservation law.
\\
The first set of conserved currents was discovered by Prasad et al
\cite{prasad}, \cite{pohlmeyer}. The construction reads as follows.
Let us rewrite the SDYM equation as
\begin{equation}
(v_y^{(n)})_{\bar{y}} + (v_z^{(n)})_{\bar{z}}=0\;\;\; \mbox{and}
\;\;\; v^{(1)}_y=J^{-1}J_y, \; v^{(1)}_z=J^{-1}J_z,
\end{equation}
where $v^{(n)}_y, v^{(n)}_y, n=1,2,3...$ are higher conserved
currents, which can be constructed by induction (iteratively). One
has to define a set of potentials $X^{(n)}$
\begin{equation}
v^{(n)}_y=\partial_{\bar{z}}X^{(n)}, \;\;\;
v^{(n)}_z=-\partial_{\bar{y}}X^{(n)}, \;\;\; X^{(0)}=I.
\end{equation}
Then, if the $n$-th current has been found, the next one is given by the
formula
\begin{equation}
v^{(n+1)}_y=(\partial_y+J^{-1}J_y)X^{(n)}, \;\;\;
v^{(n+1)}_z=(\partial_z+J^{-1}J_z)X^{(n)}.
\end{equation}
A different family of nonlocal conservation laws, nontrivially
related to Prasad's ones, was presented by Papachristou
\cite{papachristou1}. The basic idea was to reformulate the SDYM
equation using the potential symmetries. At the beginning we have a
SDYM field $J$ obeying $F[J]=0$ and introduce a potential $X$
(similarly as in Prasad's work)
\begin{equation}
J^{-1}J_y \equiv X_{\bar{z}}, \;\;\;\; J^{-1}J_z \equiv
-X_{\bar{y}}. \label{def potential}
\end{equation}
The consistency (integrability) condition
$(X_{\bar{z}})_{\bar{y}}=(X_{\bar{y}})_{\bar{z}}$ gives $F[J]=0$,
whereas the condition $(J_y)_z=(J_z)_y$ leads to the potential SDYM
equation (PSDYM)
\begin{equation}
G[X]\equiv X_{y\bar{y}}+X_{z\bar{z}}-[X_{\bar{y}},X_{\bar{z}}]=0.
\label{psdym}
\end{equation}
The point is that this expression may  also be written as a
conservation law
\begin{equation}
\partial_{\bar{y}} \left( X_y-\frac{1}{2}[X,X_{\bar{z}}]\right) + 
\partial_{\bar{z}}
\left( X_z+\frac{1}{2}[X,X_{\bar{y}}]\right)=0.
\end{equation}
Therefore, we arrive at a new current. This procedure may be
repeated. We introduce a new potential $Y$ to the last formula
\begin{equation}
X_y-\frac{1}{2}[X,X_{\bar{z}}]=Y_{\bar{z}}, \;\;\;\;
X_z+\frac{1}{2}[X,X_{\bar{y}}]=Y_{\bar{y}}
\end{equation}
and consider the consistency condition $ (X_y)_z=(X_z)_y$. As a
result we derive a new PSDYM equation which has the form of a
conservation law, as well. One may continue with this procedure 
and, at least in principle, derive an infinite set of conserved
quantities. There is some similarity between the two sets of currents,
however, the relation between them is non-trivial
\cite{papachristou1}.
\\
The importance of the PSDYM equation originates in the observation
that there is a one-to-one correspondence between symmetries of the
SDYM and PSDYM, as it was formulated in the theorem by Papachristou
\cite{papachristou2}
\begin{equation}
\delta X=\alpha \Phi\;\; \mbox{is a symmetry of} \;\; G[X]
\Leftrightarrow \delta J=\alpha J\Phi \;\; \mbox{is a symmetry
of}\;\; F[J],
\end{equation}
where $X \rightarrow X'=X+\alpha \Phi$ is a transformation which
leaves the PSDYM invariant: $G[X']=0$ if $G[X]=0$, or in other words
\begin{equation}
\delta G \equiv
H(\Phi)=\Phi_{y\bar{y}}+\Phi_{z\bar{z}}+[X_{\bar{z}},\Phi_{\bar{y}}]-
[X_{\bar{y}},\Phi_{\bar{z}}]=0.
\end{equation}
The next step is to find a B\"{a}cklund transformation generating
the symmetries of the PSDYM \cite{papachristou2}
\begin{equation}
\lambda \Phi'_{\bar{z}}=\Phi_y+[X_{\bar{z}},\Phi], \;\;\;\; \lambda
\Phi'_{\bar{y}}=-\Phi_z+[X_{\bar{y}},\Phi], \label{PSDYM back}
\end{equation}
provided $X$ is any given solution of the PSDYM equation, for
example (\ref{def potential}). Then starting with any local symmetry
of the PSDYM $\Phi^{(0)}$ (or the SDYM as they are in one-to-one
correspondence) one is able to construct an infinite tower of
symmetries $\{\Phi^{(n)}\}_{n=0}^{\infty}$. Moreover, as the
B\"{a}cklund transformation (\ref{PSDYM back}) immediately provides
a conservation law we get an infinite series of the conserved
quantities, each based on a particular local symmetry of the SDYM
equations.
\\
The extensive analysis of such families of conserved
quantities has been performed by Papachristou \cite{papachristou3}.
He introduced a recursion operator $\hat{R}$ which transforms one
symmetry of the PSDYM equation into  another one and is given by a
formal integration of the B\"{a}cklund transformation (\ref{PSDYM
back})
\begin{equation}
\hat{R}\equiv \partial_{\bar z}^{-1} (\partial_y +[X_{\bar{z}}, ]).
\end{equation}
To be precise, he constructed an infinite set of Lie derivatives
$\Delta^{(n)} X= \Phi^{(n)}$, where $\Phi^{(n)}$ is a symmetry of
the PSDYM equation as
\begin{equation}
\Delta^{(n)}_k X = R^{(n)} L_k X.
\end{equation}
Here $L_k$ is a symmetry operator for the PSDYM equation
corresponding to a given {\em local} symmetry. The results may be
summarized as follow.
\\
For internal symmetries $\Phi \equiv \Delta_k X \equiv L_k X =
[X,T_k]$, where $T_k$ is a basis for the $su(2)$ Lie algebra of the
gauge fields, we get that the infinite set of transformations
\begin{equation}
\Delta^{(n)}_k X = R^{(n)} L_k X = R^{(n)}[X,T_k]
\end{equation}
obeys the Kac-Moody algebra
\begin{equation}
[\;\Delta^{(m)}_i, \Delta^{(n)}_j]X=C^k_{ij}\Delta^{(m+n)}_k X.
\end{equation}
Once again, it is exactly the hidden symmetry of SDYM found by L.
Dolan.
\\
Finally let us discuss the nine local (point) symmetries of the SDYM
\begin{equation}
L_1=\partial_y, L_2=\partial_z,
L_3=z\partial_y-{\bar{y}}\partial_{\bar{z}}, L_4=y\partial_z-
{\bar{z}}\partial_{\bar{y}},L_5=y\partial_y-z\partial_z-
\bar{y}\partial_y+\bar{z}\partial_z,
\end{equation}
\begin{equation}
L_6=1+y\partial_y+z\partial_z,
L_7=1-\bar{y}\partial_y-\bar{z}\partial_z,
L_8=yL_6+\bar{z}(y\partial_{\bar{z}}-z\partial_{\bar{y}}),
L_9=zL_6+\bar{y}(z\partial_{\bar{y}}-y\partial_{\bar{z}})
\end{equation}
The subset $\{L_1...L_5\}$ provides an infinite set of
transformations
\begin{equation}
\Delta^{(n)}_k X = R^{(n)} L_k X, k=1..5
\end{equation}
satisfying also a Kac-Moody algebra. Additionally, $L_6$ and $L_7$ give
two sets of infinitely many transformations
\begin{equation}
\Delta^{(n)} X = R^{(n)} L X, \;\;\; L=L_6 \; \mbox{ or } \; L_7
\end{equation}
leading to two copies of the Virasoro algebra. Generators $L_8$ and
$L_9$ probably do not result in any algebraic structure.
\\
Obviously, conservation laws do not have to correspond to conserved
charges. This happens, e.g.,  if the spatial integrals of the fluxes 
(charge densities) do
not converge. As observed by Ioannidou and Ward \cite{ward}, the nonlocal
currents found by Prasad \cite{prasad} and Papachristou
\cite{papachristou1}, \cite{papachristou2} lead to densities which
diverge after integration. To be precise, it was discussed for the
chiral model in (2+1) dimension but these results should hold also for SDYM. A
general argument is the following. All nonlocal conserved currents
of type \cite{prasad}, \cite{papachristou1}, \cite{papachristou2},
\cite{papachristou3} are constructed using the integral operator
$\partial^{-1}$ and, further, the instanton field is power-like localized. Thus,
after a sufficient number of integrations we arrive at a divergent
quantity.
\\
\section{Generalized integrability in the SDYM}
Here, following \cite{joaquin1}, we very briefly describe the
self-dual sector of $SU(2)$ Yang-Mills theory in the language of
generalized integrability. The basic step in this framework is the
choice of a reducible Lie algebra $\tilde{\mathcal{G}}=\mathcal{G}
\oplus \mathcal{H}$, where $\mathcal{G}$ is a Lie algebra and
$\mathcal{H}$ is an Abelian ideal (in practice, a representation
space of $\mathcal{G}$), together with a connection
$\mathcal{A}_{\mu} \in \mathcal{G}$ and a vector field
$\mathcal{B}_\mu \in \mathcal{H}$. A system possesses the
generalized zero curvature representation if its equations of motion
may be encoded in two conditions. Namely, the flatness of the
connection
\begin{equation}
\partial_{\mu} \mathcal{A}_{\nu} - \partial_{\nu}
\mathcal{A}_{\mu}+ [\mathcal{A}_{\mu}, \mathcal{A}_{\nu}]=0
\end{equation}
and the covariant constancy of the vector field
\begin{equation}
\partial_{\mu} \mathcal{B}^{\mu} +[\mathcal{A}_{\mu},
\mathcal{B}^{\mu}]=0.
\end{equation}
Usually, one assumes a trivial connection i.e.,
$A_{\mu}=g^{-1}\partial_{\mu}g$, where $g \in G$. In this case, one
can easily construct conserved currents
$$
\mathcal{J}_{\mu}=g \mathcal{B}_{\mu} g^{-1}.
$$
We say that a system is integrable if the number of currents is
infinite. As it is equal to the dimension of the Abelian ideal
$\mathcal{H}$, the integrability condition is simply $\mbox{dim} \;
\mathcal{H} = \infty$.
\\
Let us now express the self-dual equations of $SU(2)$ YM in this
manner. Again, we use the representation of the self-dual equations
via the equation for the $J$ matrix. In order to accomplish that we
introduce a flat connection $\mathcal{A}_{\mu}$ and a covariantly
constant vector $\mathcal{B}_{\mu}$ taking values in an Abelian
ideal in the following way
\begin{equation}
\mathcal{A}_{\mu}=J^{-1} \partial_{\mu} J
=\mathcal{A}_{\mu}^r \mathcal{T}_r
\end{equation}
\begin{equation}
\mathcal{B}_{\bar{y}}= \mathcal{A}_{\bar{y}}^r \mathcal{S}_r, \;\;\;
\mathcal{B}_{\bar{z}}= \mathcal{A}_{\bar{z}}^r \mathcal{S}_r, \;\;\;
\mathcal{B}_y=0, \;\;\; \mathcal{B}_z=0,
\end{equation}
where $\mathcal{T}_r$, $\mathcal{S}_r$ form a basis satisfying
$$[\mathcal{T}_r,\mathcal{T}_s]=C^u_{rs} \mathcal{T}_u, \;\;\;
[\mathcal{T}_r, \mathcal{S}_s]=C^u_{rs} \mathcal{S}_u, \;\;\;
[\mathcal{S}_r,\mathcal{S}_s]=0.
$$
Obviously, the connection is flat as it is a pure gauge
configuration. Moreover the condition for the vector field i.e.,
$D_{\mu} \mathcal{B}^{\mu}=0$ is equivalent to the self-dual
equation (\ref{sdym J}). One can construct conserved currents
\begin{equation}
\mathcal{J}_{\bar{y}}= \mathcal{A}^r_{\bar{y}} J
\mathcal{S}_r J^{-1} \;\;\; \mathcal{J}_{\bar{z}}=
\mathcal{A}^r_{\bar{z}} J \mathcal{S}_r J^{-1},
\;\;\; \mathcal{J}_y=0, \;\;\; \mathcal{J}_z=0 ,
\end{equation}
then, the conservation laws are just the self-dual equations
(\ref{sdym J}). More conservation laws may be derived as discussed in
the previous section.
\\
Of course, the obtained result is not surprising. A system which
possesses the standard zero curvature representation admits also the
generalized zero curvature formulation. However, there is a simple
prescription how to construct an infinite family of additional conserved
currents for a model with generalized zero curvature formulation. In
general, they are spanned by the canonical momenta conjugated to
the field degrees of freedom. It is important to check whether such
currents can be also found for the self-dual sector of the $SU(2)$
YM theory, and, if the answer is positive, what is their relation
with the standard non-local conservation laws described before.
\section{New conserved currents in the SDYM}
\subsection{Cho--Faddeev--Niemi--Shabanov decomposition}
In order to derive such conserved quantities in an exact form we
perform a nonlocal change of variables known as the
Cho--Faddeev--Niemi--Shabanov decomposition \cite{cho1} -
\cite{gies}. The decomposition
\begin{equation}
\vec{A}_\mu =C_{\mu}\vec{n}+\partial_{\mu}\vec{n} \times \vec{n} + \vec
W_{\mu} \label{decomp}
\end{equation}
relates the original $SU(2)$ non-Abelian gauge field with three
fields: a three component unit vector field $\vec{n}$ pointing into
the color direction, an Abelian gauge potential $C_{\mu}$ and a
color vector field $W_{\mu}^a$ which is perpendicular to $\vec{n}$.
The fields are not independent. In fact, as we want to keep the
correct gauge transformation properties
\begin{equation}
\delta n^a=\epsilon^{abc}n^b\alpha^c, \;\;\; \delta
W^a_{\mu}=\epsilon^{abc}W^b_{\mu}\alpha^c, \;\;\; \delta
C_{\mu}=n^a\alpha^a_{\mu} \label{gauge trans decomp}
\end{equation}
under the primary gauge transformation
\begin{equation}
\delta A^a_{\mu}=(D_{\mu} \alpha)^a = \alpha^a_{\mu} +
\epsilon^{abc} A^b_{\mu} \alpha^c \label{gauge trans su2}
\end{equation}
one has to impose the constraint ($n^b_{\mu}\equiv \partial_\mu n^b$
etc.)
\begin{equation}
\partial^{\mu}W^a_{\mu}+C^{\mu}\epsilon^{abc}n^bW^c_{\mu} + n^a
W^{b\mu}n^b_{\mu}=0. \label{decomp constrain}
\end{equation}
In the subsequent analysis we assume a particular form for the
valence field $W_{\mu}^a$. It is equivalent to a partial gauge
fixing where one leaves a residual local $U(1)$ gauge symmetry.
Namely,
\begin{equation} \label{rho-sig-eq}
W^a_{\mu} =  \rho
 n^a_{\mu} + \sigma \epsilon^{abc} n^b_{\mu} n^c, \label{W def}
\end{equation}
where $\rho, \sigma$ are real scalars. For reasons of convenience we
combine them into a complex scalar $v=\rho + i \sigma$. Then the
Lagrange density takes the form ($u_\mu \equiv \partial_\mu u$ etc.)
\begin{eqnarray}
L=F^2_{\mu \nu}  - 2(1-|v|^2)H_{\mu\nu} + (1-|v|^2)^2 H_{\mu\nu}^2 +
\frac{8}{(1+|u|^2)^2} \left[ (u_{\mu} \bar{u}^{\mu}) (D^{\nu} v
\overline{D_{\nu}v}) - (D_{\mu}v \bar{u}^{\mu}) (\overline{D_{\nu}v}
u^{\nu}) \right], \label{lagrangian}
\end{eqnarray}
where
\begin{equation}
H_{\mu \nu}= \vec{n} \cdot \left[ \vec{n}_{\mu} \times \vec{n}_{\nu}
\right]= \frac{-2i}{(1+|u|^2)^2} (u_{\mu}\bar{u}_{\nu} - u_{\nu}
\bar{u}_{\mu} ), \;\;\;\;\; H_{\mu \nu}^2= \frac{8}{(1+|u|^2)^4} [
(u_{\mu} \bar{u}^{\mu})^2 -u_{\mu}^2\bar{u}^2_{\nu} ] \label{H}
\end{equation}
and the covariant derivatives read $D_{\mu} v =v_{\mu} -ieC_{\mu}v$,
$\overline{D_{\mu} v}=\bar{v}_{\mu}+ieC_{\mu}\bar{v}$ and we
expressed the unit vector field by means of the stereographic
projection
$$
\vec{n}=\frac{1}{1+|u|^2} \left(u+\bar{u},-i(u-\bar{u}), |u|^2-1
\right).
$$
Further,
$$
F_{\mu \nu}\equiv \partial_\mu C_\nu -\partial_\nu C_\mu
$$ 
is the Abelian field strength tensor corresponding to
the Abelian gauge field $C_{\mu}$. Notice that only the complex
field $v$ couples to the gauge field via the covariant derivative.
\subsection{Self-dual equations}
Now, we apply the CFNS decomposition to the self-dual equations. As
we know the full field strength tensor reads
\begin{eqnarray}
\vec{F}_{\mu \nu}= \left[ F_{\mu \nu} - (1-|v|^2)H_{\mu \nu} \right]
\vec{n} + \frac{1}{2} \left[(D_{\mu} v + \overline{D_{\mu} v})
\vec{n}_{\nu} - (D_{\nu} v + \overline{D_{\nu} v} )\vec{n}_{\mu}
\right] + \nonumber
\\ \frac{1}{2i} \left[(D_{\mu} v - \overline{D_{\mu} v})
\vec{n}_{\nu} \times \vec{n} - (D_{\nu} v - \overline{D_{\nu} v})
\vec{n}_{\mu}\times \vec{n} \right]. \label{sd F decomp}
\end{eqnarray}
Therefore, using the self-dual equations (\ref{sd def}) we get two
expressions, one parallel and one perpendicular to the color vector
$\vec{n}$
\begin{equation}
\frac{1}{2} \epsilon_{\mu \nu \rho \sigma} [F^{\rho \sigma} -
(1-|v|^2)H^{\rho \sigma}]=F_{\mu \nu} - (1-|v|^2)H_{\mu \nu}
\label{sd tang}
\end{equation}
and
\begin{eqnarray}
\frac{1}{2} \epsilon_{\mu \nu }{}^{\rho \sigma} \left[ \left[(D_{\rho} v
+ \overline{D_{\rho} v}) \vec{n}_{\sigma} - (D_{\sigma} v +
\overline{D_{\sigma} v} )\vec{n}_{\rho} \right]  -i \left[(D_{\rho}
v - \overline{D_{\rho} v}) \vec{n}_{\sigma} \times \vec{n} -
(D_{\sigma} v - \overline{D_{\sigma} v}) \vec{n}_{\rho}\times
\vec{n} \right] \right]= \nonumber
\\ \left[(D_{\mu} v + \overline{D_{\mu}
v}) \vec{n}_{\nu} - (D_{\nu} v + \overline{D_{\nu} v} )\vec{n}_{\mu}
\right] + i \left[(D_{\mu} v - \overline{D_{\mu} v}) \vec{n}_{\nu}
\times \vec{n} - (D_{\nu} v - \overline{D_{\nu} v})
\vec{n}_{\mu}\times \vec{n} \right]. \label{sd perp}
\end{eqnarray}
For later convenience we now want to derive some constraints which
result from these two sets of equations.
On the one hand, after projection on $\vec{n}_{\mu}$, Eq. (\ref{sd
perp}) gives
\begin{eqnarray}
(D_{\mu} v +\overline{D_{\mu} v}) \vec{n}^{\mu} \cdot \vec{n}_{\nu}
- (D_{\nu} v +\overline{D_{\nu} v}) \; \vec{n}^2_{\mu} -i(D^{\mu} v
- \overline{D^{\mu} v}) H_{\mu \nu}= -i \epsilon^{\mu \nu \lambda
\omega} (D_{\lambda} v - \overline{D_{\lambda} v}) H_{\mu \omega}.
\label{sd perp1}
\end{eqnarray}
On the other hand, if we multiply (\ref{sd perp}) by $\times
\vec{n}_{\mu}$ and project on $\vec{n}$ then we get
\begin{eqnarray}
(D_{\mu} v -\overline{D_{\mu} v}) \vec{n}^{\mu} \cdot \vec{n}_{\nu}
- (D_{\nu} v -\overline{D_{\nu} v}) \; \vec{n}^2_{\mu} -i(D^{\mu} v
+\overline{D^{\mu} v}) H_{\mu \nu}= -i \epsilon^{\mu \nu \lambda
\omega} (D_{\lambda} v + \overline{D_{\lambda} v}) H_{\mu \omega}.
\label{sd perp2}
\end{eqnarray}
Both equations lead to the simple expression
\begin{eqnarray}
D_{\nu} v (u_{\mu} \bar{u}^{\mu}) - u_{\nu} (D_{\mu} v
\bar{u}^{\mu}) = \epsilon_{\mu \nu \rho \sigma} D^{\rho} v u^{\mu}
\bar{u}^{\sigma} \label{sd perp gen}
\end{eqnarray}
and its complex conjugate. This expression just constitutes a system
of linear homogeneous algebraic equations for the unknowns
$D^{\mu}v$,
\begin{eqnarray}
M_{\mu \nu} D^{\mu}v=0, \;\;\;\;\; M_{\mu \nu}= (u_{\alpha}
\bar{u}^{\alpha}) \delta_{\mu \nu} - u_{\mu}\bar{u}_{\nu} -
\epsilon_{\mu \nu \rho \sigma}u^{\rho}\bar{u}^{\sigma}. \label{sd
set}
\end{eqnarray}
In order to find all solutions of this system of equations we
consider the corresponding eigenvalue problem $M_{\mu \nu}
D^{\mu}v=\lambda D_{\mu} v$. Of course, a solution exists if and
only if the determinant vanishes
\begin{equation}
\mbox{Det} (\hat{M} - \lambda I)=0.
\end{equation}
On the other hand one can find that
\begin{equation}
\mbox{Det} (\hat{M} - \lambda I)=\lambda (\lambda -u_{\mu}
\bar{u}^{\mu})(u_{\mu}^2 \bar{u}_{\nu}^2 - 2 \lambda u_{\mu}
\bar{u}^{\mu} + \lambda^2).
\end{equation}
Generically there is a single eigenvalue  $\lambda=0$ corresponding
to the solution
\begin{equation}
D_{\mu}v= f u_{\mu},
\end{equation}
where $f$ is an arbitrary function. However, if the complex field
$u$ obeys the complex eikonal equation $u_{\mu}^2=0$, then $\lambda
=0 $ is a degenerate eigenvalue with degeneracy 2. In this case,
there exists a second solution. This second solution may be
expressed more easily in terms of real vectors. Indeed, if we write
$u=a+ib$ then the complex eikonal equation corresponds to
\begin{equation}
a^\mu b_\mu =0 \, ,\quad a_\mu^2 = b_\mu^2 .
\end{equation}
If we introduce analogously $D_\mu v = c_\mu +i d_\mu$ then the
second solution is given by
\begin{equation} \label{sec-sol-cond1}
a^\mu c_\mu = b^\mu c_\mu = a^\mu d_\mu = b^\mu d_\mu =c^\mu d_\mu
=0
\end{equation}
and
\begin{equation} \label{sec-sol-cond2}
c_\mu^2 = d_\mu^2 .
\end{equation}
The vector $D_\mu v = c_\mu +id_\mu$ is unique up to a
multiplication by an arbitrary complex function, as befits the
solution to a complex, homogeneous linear equation. Conditions
(\ref{sec-sol-cond1}), (\ref{sec-sol-cond2}) imply that the complex
vector $D_\mu v$ has to obey
\begin{equation}
u^\mu D_\mu v = \bar u^\mu D_\mu v =0 \, ,\quad D^\mu v  D_\mu v =0
\end{equation}
in order to be a solution of the second type.
We remark that a wide class of explicitly known instanton
configurations, like, e.g., the cylindrically symmetric solutions
found by Witten \cite{Witt1}, belongs to this second case.
\\
A further possibility, $u_{\alpha} \bar{u}^{\alpha}=0$, which would
lead to a even higher degeneracy, is physically uninteresting since
it leads to the trivial solutions $u=$ const.
\\
Taking into account formula (\ref{sd perp gen}) and its general
solutions discussed above we find three constraints which are
satisfied by all self-dual configurations
\begin{equation}
(D_{\lambda}v u^{\lambda})(u^{\beta}\bar{u}_{\beta}) -(D_{\lambda}v
\bar{u}^{\lambda}) u_{\beta}^2 =0, \label{sd c1}
\end{equation}
\begin{equation}
(D_{\lambda}v)^2 (u^{\beta}\bar{u}_{\beta}) -(D_{\lambda}v
\bar{u}^{\lambda}) (D^{\beta} v u_{\beta}) =0, \label{sd c2}
\end{equation}
\begin{equation}
(D^{\nu} v \overline{D_{\nu}v}) u_{\mu}^2- (\overline{D_{\nu}v
}u^{\nu}) (D_{\mu}v u^{\mu})=0. \label{sd c23}
\end{equation}
\subsection{Conserved currents}
Following considerations presented, e.g.,  in \cite{ASG2} -
\cite{ASGW2}, the family of conserved currents may be constructed in the
following form
\begin{equation}
j_{\mu}^{G}= i(1+|u|^2)^2 \left(\bar{\pi}_{\mu} \frac{\partial
G}{\partial u}- \pi_{\mu} \frac{\partial G}{\partial \bar{u}}
\right), \label{curr def}
\end{equation}
where $G$ is an arbitrary real function of the complex field $u$ i.e.,
$G=G(u,\bar{u})$ and $\pi_{\mu}$ is the canonical momentum (\ref{pi
def}). The four-divergence reads ($G_u \equiv \partial_u G$ etc.)
\begin{eqnarray}
 \partial^{\mu} j_{\mu}^G= i(1+|u|^2)^2 \left[ G_u\partial_{\mu}
\bar{\pi}^{\mu}-G_{\bar{u}}
 \partial_{\mu} \pi^{\mu} +G_{uu}u_{\mu}\bar{\pi}^{\mu}
 +G_{u\bar{u}} \bar{u}_{\mu}\bar{\pi}^{\mu}  -G_{\bar{u}u} u_{\mu}
 \pi^{\mu} -G_{\bar{u} \bar{u}} \bar{u}_{\mu} \pi^{\mu} \right]+
\nonumber \\
2i(1+|u|^2)(u\bar{u}_{\mu}+\bar{u}u_{\mu}) (G_u
\bar{\pi}^{\mu}-G_{\bar{u}} \pi^{\mu}). \label{4 div}
\end{eqnarray}
or
\begin{eqnarray}
 \partial^{\mu} j_{\mu}^G= i(1+|u|^2)^2 \left[ G_u
\left( \partial_{\mu} \bar{\pi}^{\mu}+
 \frac{2u}{1+|u|^2}\bar{u}_{\mu}\bar{\pi}^{\mu}\right) -
G_{\bar{u}} \left(\partial_{\mu} \pi^{\mu} +
 \frac{2\bar{u}}{1+|u|^2} \pi_{\mu} u^{\mu} \right) +
G_{u\bar{u}} (\bar{u}_{\mu}\bar{\pi}^{\mu} - u_{\mu}
 \pi^{\mu}) \right] \nonumber \\  + i(1+|u|^2)^2
\left[\left(G_{uu} +\frac{2\bar{u}G_u}{1+|u|^2}
 \right) u_{\mu}\bar{\pi}^{\mu} -
\left(G_{\bar{u} \bar{u}} + \frac{2uG_{\bar{u}}}{1+|u|^2}\right)
\bar{u}_{\mu} \pi^{\mu} \right]. \label{4 div a}
\end{eqnarray}
Taking into account that $\bar{u}_{\mu}\bar{\pi}^{\mu}=u_{\mu}
 \pi^{\mu}$ and the pertinent field equations $ (1+|u|^2)
\partial_{\mu} \pi^{\mu}
 + 2 \bar{u}\pi_{\mu} u^{\mu}=0$ we get
\begin{eqnarray}
 \partial^{\mu} j_{\mu}^G= i(1+|u|^2)^2
\left[\left(G_{uu} +\frac{2\bar{u}G_u}{1+|u|^2}
 \right) u_{\mu}\bar{\pi}^{\mu} -
\left(G_{\bar{u} \bar{u}} + \frac{2uG_{\bar{u}}}{1+|u|^2}\right)
\bar{u}_{\mu} \pi^{\mu} \right]. \label{4 div b}
\end{eqnarray}
Due to the arbitrariness of the function $G$ the currents are
conserved if $u_{\mu}\bar{\pi}^{\mu}=0$ and $\bar{u}_{\mu}
\pi^{\mu}=0$. These so-called integrability conditions introduce
some new relations between degrees of freedom and, in principle, do
not have to be satisfied for all solutions of Yang-Mills theory.
However, it turns out that in the self-dual sector both conditions
hold identically. To prove it observe that
\begin{eqnarray}
\bar{u}_{\mu} \pi^{\mu}= \frac{8}{(1+|u|^2)^2} \left[  (D^{\nu} v
\overline{D_{\nu}v}) \bar{u}_{\mu}^2- (D_{\nu}v \bar{u}^{\nu})
\overline{D_{\mu}v} \bar{u}^{\mu}
 \right],
\end{eqnarray}
where we have used the antisymmetry of $F_{\mu \nu}$ and $K_{\mu}
\bar{u}^{\mu} \equiv 0$ (where $K_\mu$ is defined in (\ref{K-vec})). 
The resulting expression is just the
complex conjugate of formula (\ref{sd c23}) and therefore equals
zero for all configurations of the self-dual sector.

The charges corresponding to the currents (\ref{curr def}) are
\begin{equation}
Q^G \equiv \int d^3 x j_0^G
\end{equation}
obey the algebra of area-preseving diffeomorphisms on the target space
two-sphere spanned by the field $u$ under the Poisson bracket,
where the fundamental Poisson bracket is (with $x^0 =y^0$)
\begin{equation}
\{ u({\bf x}),\pi ({\bf y}) \} = \{ \bar u({\bf x}),\bar \pi ({\bf y}) \}   
=\delta^3 ({\bf x} - {\bf y}) ,
\end{equation}
as usual. Explicitly, the algebra of area-preserving diffeomorphisms is
\begin{equation}
\{ Q^{G_1}, Q^{G_2} \} = Q^{G_3} \quad ,
\qquad G_3 = i (1+|u|^2)^2
(G_{1,\bar u} G_{2,u} - G_{1,u}G_{2,\bar u}) .
\end{equation}
Finally, let us remark that the currents (\ref{curr def}) are 
invariant under the residual
U(1) gauge transformations that remain after the partial gauge fixing 
implied by the CFNS decomposition, see Eq. (\ref{rho-sig-eq}).

\subsection{Trivially conserved currents}
Using this method we are able to construct more families of
infinitely many conserved quantities in self-dual Yang--Mills
theory, which are based on other canonical momenta. They are given
by the expressions
\begin{equation}
j_{\mu}^{H}= \bar{P}_{\mu}\frac{\partial H}{\partial v}-P_{\mu}
\frac{\partial H}{\partial \bar{v}}, \label{med cur H def}
\end{equation}
\begin{equation}
j_{\mu}^{\tilde{G}}= \omega_{\mu \nu} \left( \frac{\partial
\tilde{G}}{\partial u} u^{\nu}+ \frac{\partial \tilde{G}}{\partial
\bar{u}} \bar{u}^{\nu} \right), \label{weak cur 3}
\end{equation}
\begin{equation}
j_{\mu}^{\tilde{H}}=\omega_{\mu \nu} \left( \frac{\partial
\tilde{H}}{\partial v} D^{\nu}v +\frac{\partial \tilde{H}}{\partial
\bar{v}} \overline{D^{\nu}v} \right), \label{weak cur 4}
\end{equation}
where the function $H=H(u,\bar{u},v\bar{v})$ while the functions
$\tilde{G}, \tilde{H}$ depend on the moduli only
$\tilde{G}=\tilde{G}(u\bar{u},v\bar{v}),
\tilde{H}=\tilde{H}(u\bar{u},v\bar{v})$. However, all these currents
are trivially conserved. To see this let us analyze the first family
in detail. First of all observe that it may be written as
\begin{equation}
j_{\mu}^{H}= H'(\bar{v} \bar{P}_{\mu}- v P_{\mu}) \label{med cur H
def 1}
\end{equation}
where the prime denotes the derivative w.r.t. $v\bar v$, and $P_\mu$ is
defined in Eq. (\ref{P-vec}) . Using the
self-dual equations we find that
\begin{equation}
j_{\mu}^{H}= \frac{8H'}{(1+|u|^2)^2} \left(\epsilon_{\alpha \mu
\beta \gamma} u^{\alpha}(\bar{v} D^{\beta}v + v\overline{D^{\beta}v})
\bar{u}^{\gamma} \right)=\frac{8H'}{(1+|u|^2)^2}
\left(\epsilon_{\alpha \mu \beta \gamma} u^{\alpha}(\bar{v} v^{\beta}
+ v\bar{v}^{\beta}) \bar{u}^{\gamma} \right). \label{med cur H def 2}
\end{equation}
Therefore, these currents are conserved entirely due to the
antisymmetry of the $\epsilon_{\alpha \mu \beta \gamma}$ tensor.
Analogously one can check that the two remaining families are
trivially conserved, as well.
\section{Conclusions}
The main achievement of the present paper is the derivation of a new
family of infinitely many conserved currents for the self-dual
sector of classical $SU(2)$ YM theory. This has been accomplished by
a combination of techniques developed in the so-called
generalized integrability (generalized zero curvature) formulation
with a nonlocal transformation of the original gauge degrees of
freedom (CFNS decomposition). This alternative procedure provides
currents with rather different properties than the previously known
ones.
\\
First of all, all calculations are done in a completely covariant
manner. Therefore, the obtained currents are conserved for the self-dual
sector of $SU(2)$ YM in space-times in any dimension with a completely
arbitrary signature.
\\
Secondly, these new currents have a more standard geometrical origin.
They are the Noether currents corresponding to the area preserving
diffeomorphisms on the two dimensional target space. Therefore they
obey the classical diffeomorphism algebra instead of the Kac-Moody
or Virasoro ones. Also, the relation between conservation laws and symmetries
is different in our case. Although the currents we found generate 
area-preserving diffeomorphisms on target space, this does not imply that
these diffeomorphisms are symmetries of the SDYM equations. The reason is that
the SDYM equations in the CFNS decomposition are not Euler--Lagrange,
therefore the Noether theorem does not apply (observe that the canonical
momenta are derived from the Lagrangian of the original Yang--Mills system,
which gives rise to the full Yang--Mills equations).
\\
Thirdly, the currents derived here are given in an explicit form. This
is an advantage in comparison with the currents of Prasad and
Papachristou, which are given in a more 
complicated, iterative way and are, therefore, not so easy to work with.
\\
Finally, let us briefly mention some possible generalizations and further
directions of future investigations. On the one hand, the procedure employed
here
is based on the generalized zero curvature condition of Ref. \cite{joaquin1}, 
which is not restricted to the SDYM. It has been and will be used to detect
further integrable sectors in different field theories. On the other hand,
recently other nonlocal decompositions of Yang--Mills theory have been
proposed, like, e.g., the spin-charge separation of \cite{FN1} - \cite{Fad1}.
It is an interesting question whether these decompositions allow to detect 
further conservation laws in SDYM. This problem is under current investigation.

\section*{Acknowledgements}

C.A. and J.S.-G. thank MCyT (Spain) and FEDER (FPA2005-01963), and
support from
 Xunta de Galicia (grant PGIDIT06PXIB296182PR and Conselleria de
Educacion). A.W. acknowledges support from the Foundation for Polish
Science FNP (KOLUMB programme) and Ministry of Science and Higher
Education of Poland (grant N N202 126735).

\appendix
\section*{Appendix}
\setcounter{section}{1} Here we calculate the canonical momenta
\begin{eqnarray}
\pi_{\mu} =\frac{\partial L}{\partial u^{\mu}}= 8i
\frac{1-|v|^2}{(1+|u|^2)^2} F_{\mu\nu} \bar{u}^{\nu} + 16
\frac{(1-|v|^2)^2}{(1+|u|^2)^4} K_{\mu} + \frac{8}{(1+|u|^2)^2}
\left[  (D^{\nu} v \overline{D_{\nu}v}) \bar{u}_{\mu}- (D^{\nu}v
\bar{u}_{\nu}) \overline{D_{\mu}v}
 \right], \label{pi def}
\end{eqnarray}
\begin{eqnarray}
\bar{\pi}_{\mu}= \frac{\partial L}{\partial \bar{u}^{\mu}} = -8i
\frac{1-|v|^2}{(1+|u|^2)^2} F_{\mu\nu} u^{\nu} + 16
\frac{(1-|v|^2)^2}{(1+|u|^2)^4} \bar{K}_{\mu} +
 \frac{8}{(1+|u|^2)^2} \left[  (D^{\nu} v \overline{D_{\nu}v})
u_{\mu}- (\overline{D^{\nu}v }u_{\nu}) D_{\mu}v
 \right] \label{pi b def}
\end{eqnarray}
where
\begin{equation} \label{K-vec}
K_{\mu}=(u_{\nu} \bar{u}^{\nu}) \bar{u}_{\mu} - \bar{u}_{\nu}^2 u_{\mu}
\end{equation}
and
\begin{equation} \label{P-vec}
P_{\mu}=\frac{\partial L}{\partial v^{\mu}} =\frac{8}{(1+|u|^2)^2}
\left[ (u_{\nu} \bar{u}^{\nu})\overline{D_{\mu}v} -
(\overline{D^{\nu}v }u_{\nu}) \bar{u}_{\mu} \right], \label{P def}
\end{equation}
\begin{equation}
\bar{P}_{\mu}=\frac{\partial L}{\partial \bar{v}^{\mu}}=
\frac{8}{(1+|u|^2)^2}\left[ (u_{\nu} \bar{u}^{\nu})D_{\mu}v -
(D^{\nu}v \bar{u}_{\nu}) u_{\mu} \right] \label{P b def}
\end{equation}
and finally
\begin{equation}
\omega_{\mu \nu}=\frac{\partial L}{\partial
(\partial^{\mu}C^{\nu})}=4 \left(F_{\mu \nu} - (1-|v|^2)H_{\mu \nu}
\right). \label{omega def}
\end{equation}
The pertinent equations of motion for the complex $u$ field read
\begin{eqnarray}
\partial_{\mu} \pi^{\mu}=L_u= -16 i \bar{u}
\frac{1-|v|^2}{(1+|u|^2)^3} F^{\mu\nu} u_{\mu} \bar{u}_{\nu} -
4\cdot 8 \bar{u} \frac{(1-|v|^2)^2}{(1+|u|^2)^5} \left[
(u_{\mu}\bar{u}^{\mu})^2 -
u_{\mu}^2 \bar{u}^2_{\nu} \right] \nonumber \\
- \frac{16\bar{u}}{(1+|u|^2)^3} \left[ (u_{\mu} \bar{u}^{\mu})
(D^{\nu} v \overline{D_{\nu}v}) - (D^{\mu}v \bar{u}_{\mu})
(\overline{D_{\nu}v} u^{\nu}) \right] \label{eom pi}
\end{eqnarray}
\begin{eqnarray}
\partial_{\mu} \bar{\pi}^{\mu}=L_{\bar{u}}= -16 i u
\frac{1-|v|^2}{(1+|u|^2)^3} F^{\mu\nu} u_{\mu} \bar{u}_{\nu} -
4\cdot 8 u \frac{(1-|v|^2)^2}{(1+|u|^2)^5} \left[
(u_{\mu}\bar{u}^{\mu})^2 - u_{\mu}^2 \bar{u}^2_{\nu}
\right] \nonumber \\
- \frac{16u}{(1+|u|^2)^3} \left[ (u_{\mu} \bar{u}^{\mu}) (D^{\nu} v
\overline{D_{\nu}v}) - (D^{\mu}v \bar{u}_{\mu}) (\overline{D_{\nu}v}
u^{\nu}) \right], \label{eom pi b}
\end{eqnarray}
while for the complex $v$ field we get
\begin{eqnarray}
\partial_{\mu} P^{\mu}=L_{v}=\frac{-8 i\bar{v}}{(1+|u|^2)^2}
F^{\mu\nu} u_{\mu} \bar{u}_{\nu} + \frac{2\cdot
8\bar{v}(1+|v|^2)}{(1+|u|^2)^4} \left[ (u_{\mu}\bar{u}^{\mu})^2 -
u_{\mu}^2 \bar{u}^2_{\nu} \right] +\nonumber \\
\frac{-8ie}{(1+|u|^2)^2} \left[ (u_{\mu} \bar{u}^{\mu}) (C^{\nu}
\overline{D_{\nu}v}) - (C^{\mu} \bar{u}_{\mu}) (\overline{D_{\nu}v}
u^{\nu}) \right] \label{eom P}
\end{eqnarray}
\begin{eqnarray}
\partial_{\mu} \bar{P}^{\mu}=L_{\bar{v}}=\frac{-8
iv}{(1+|u|^2)^2} F^{\mu\nu} u_{\mu} \bar{u}_{\nu} + \frac{2\cdot
8v(1+|v|^2)}{(1+|u|^2)^4} \left[ (u_{\mu}\bar{u}^{\mu})^2 -
u_{\mu}^2 \bar{u}^2_{\nu}
\right] +\nonumber \\
\frac{8ie}{(1+|u|^2)^2} \left[ (u_{\mu} \bar{u}^{\mu}) (C^{\nu}
D_{\nu}v) - (C^{\mu} u_{\mu}) (D_{\nu}v u^{\nu}) \right]. \label{eom
P b}
\end{eqnarray}
The equation for the Abelian gauge field has the form
\begin{eqnarray}
\partial_{\mu} \omega^{\mu \nu}= \frac{\partial L}{\partial
C^{\nu}}= \frac{-8ie}{(1+|u|^2)^2} \left\{ (u_{\mu} \bar{u}^{\mu})
\left[ v \overline{D^{\nu}v} - \bar{v} D^{\nu} v \right] - v
\bar{u}^{\nu} (\overline{D_{\mu}v} u^{\mu}) +
\bar{v}u^{\nu}(D_{\mu}v \bar{u}^{\mu})\right\}. \label{eom C}
\end{eqnarray}

\end{document}